\documentclass[aps,prl,twocolumn,groupedaddress]{revtex4}
\usepackage{graphicx}

\newcommand{\jpsi}{\rm J/$\psi$}
\newcommand{\psip}{$\psi^\prime$}

\newcommand{\ezdc}{$E_{\rm ZDC}$}

\begin{document}

\def\cagl{$^{1}$}
\def\cern{$^{2}$}
\def\lpc{$^{3}$}
\def\heid{$^{4}$}
\def\lis{$^{5}$}
\def\llr{$^{6}$}
\def\riken{$^{7}$}
\def\suny{$^{8}$}
\def\torino{$^{9}$}
\def\ipn{$^{10}$}
\def\yer{$^{11}$}


\title{J/$\psi$ production in Indium-Indium collisions at 158 GeV/nucleon}


\author{R.~Arnaldi$^{9}$, 
K.~Banicz$^{2,4}$, 
J.~Castor$^{3}$, 
B.~Chaurand$^{6}$, 
C.~Cical\`o$^{1}$,  
A.~Colla$^{9}$, 
P.~Cortese$^{9}$,
S.~Damjanovic$^{2,4}$, 
A.~David$^{2,5}$, 
A.~de~Falco$^{1}$, 
A.~Devaux$^{3}$, 
L.~Ducroux$^{10}$, 
H.~En'yo$^{7}$,
J.~Fargeix$^{3}$,
A.~Ferretti$^{9}$, 
M.~Floris$^{1}$, 
A.~F\"orster$^{2}$,
P.~Force$^{3}$,
N.~Guettet$^{2,3}$,
A.~Guichard$^{10}$, 
H.~Gulkanian$^{11}$, 
J.~M.~Heuser$^{7}$,
M.~Keil$^{2,5}$, 
L.~Kluberg$^{2,6}$, 
C.~Louren\c{c}o$^{2}$,
J.~Lozano$^{5}$, 
F.~Manso$^{3}$, 
P.~Martins$^{2,5}$,  
A.~Masoni$^{1}$,
A.~Neves$^{5}$, 
H.~Ohnishi$^{7}$, 
C.~Oppedisano$^{9}$,
P.~Parracho$^{2}$, 
P.~Pillot$^{10}$, 
T.~Poghosyan$^{11}$,
G.~Puddu$^{1}$, 
E.~Radermacher$^{2}$,
P.~Ramalhete$^{2}$, 
P.~Rosinsky$^{2}$, 
E.~Scomparin$^{9}$,
J.~Seixas$^{5}$, 
S.~Serci$^{1}$, 
R.~Shahoyan$^{2,5}$, 
P.~Sonderegger$^{5}$,
H.~J.~Specht$^{4}$, 
R.~Tieulent$^{10}$, 
G.~Usai$^{1}$, 
R.~Veenhof$^{2,5}$,
H.~K.~W\"ohri$^{1,2,5}$\\ 
(NA60 Collaboration)
}
\affiliation{
\cagl \mbox{Universit\`a di Cagliari and INFN, Cagliari, Italy}\\
\cern \mbox{CERN, 1211 Geneva 23, Switzerland}\\
\lpc \mbox{LPC, Universit\'e Blaise Pascal and CNRS-IN2P3, Clermont-Ferrand, 
France}\\
\heid \mbox{Physikalisches~Institut~der~Universit\"{a}t Heidelberg,~Germany}\\
\lis \mbox{Instituto Superior T\'ecnico, Lisbon, Portugal}\\
\llr \mbox{LLR, Ecole Polytechnique and CNRS-IN2P3, Palaiseau, France} \\
\riken \mbox{RIKEN, Wako, Saitama, Japan}\\
\torino \mbox{Universit\`a di Torino and INFN,~Italy}\\
\ipn \mbox{IPN-Lyon, Universit\'e Claude Bernard Lyon-I and CNRS-IN2P3, Lyon, 
France}\\
\yer \mbox{YerPhI, Yerevan Physics Institute, Yerevan, Armenia}\\
}


\date{\today}

\begin{abstract}
The NA60 experiment studies muon pair production at
the CERN SPS. In this letter we report
on a precision measurement of \jpsi\ in \mbox{In-In} collisions. We have 
studied the \jpsi\ centrality distribution, and we have compared it with the 
one expected if absorption in cold nuclear matter were the only active 
suppression mechanism.
For collisions involving more than $\sim$80 participant nucleons, we find that 
an extra suppression is present. 
This result is in qualitative agreement with previous \mbox{Pb-Pb} measurements 
by the NA50 experiment, but no theoretical explanation is presently able to 
coherently describe both results.
\end{abstract}


\maketitle


The suppression of the charmonium states by color screening has been predicted 
as a signature of the phase transition from hadronic matter 
towards a Quark-Gluon Plasma~\cite{Sat86}.
Recently, it has been pointed out that the relatively loosely bound states 
$\psi'$ and $\chi_c$ should indeed melt for temperatures around the critical 
temperature $T_{\rm c}$, while the tightly bound \jpsi\ could survive, even if 
with strong medium-induced modifications, up to 
$\sim2 T_{\rm c}$~\cite{Dat04,Asa04,Sat06}.
However, the description of the evolution of a $c\overline c$ pair formed by 
gluon fusion at early times in the history of the collision, which may 
eventually lead to the formation of a bound state, is still not fully 
explained by theory.
The influence of the medium, a percolating partonic condensate~\cite{Dig04}, or 
a fully thermalized QGP~\cite{Gra04}, or even a dense hadronic 
gas~\cite{Cap05,Mai05} has been investigated in detail, but this physics topic 
is still largely data driven, and accurate experimental data are clearly needed.

At the CERN SPS, the NA38 and NA50 experiments have already studied \jpsi\ 
production in various colliding systems, including \mbox{p-A}~\cite{Ale04}, 
\mbox{S-U}~\cite{Abr99} and \mbox{Pb-Pb}~\cite{Ale05}. Proton-nucleus data 
provide an important reference, describing the expected absorption of \jpsi\ in 
cold nuclear matter. By comparing the centrality dependence of the \jpsi\ yield 
observed in nucleus-nucleus collisions to this reference, one can look for 
suppression mechanisms connected with the formation of a strongly interacting 
medium. 
In particular, NA50 has observed, in \mbox{Pb-Pb} collisions, that below a 
certain centrality threshold the \jpsi\ production is well described invoking 
nuclear absorption as the only suppression mechanism; on the contrary, above 
that threshold, an extra suppression (also known as ``anomalous'' suppression) 
sets in.
Such an interesting observation needs to be complemented by further sets of 
accurate measurements obtained with different collision systems. In this way 
one can study in more detail the onset of the anomalous suppression, and 
understand which is the physics mechanism at its origin.

The NA60 experiment has studied \jpsi\ production in \mbox{In-In} collisions at 
158 GeV/nucleon at the CERN SPS, through its decay into two muons. The 
experimental apparatus is based on a muon spectrometer (MS), inherited from 
NA50~\cite{Abr97}, used for triggering on muon pair production and for 
tracking purposes. A 12$\lambda_i$ thick hadron absorber, 
mostly made of graphite, separates the MS from the target region,
equipped with a beam tracker (BT) 
and a vertex tracker (VT), placed inside a 2.5 T dipole magnet.
Finally, a Zero Degree Calorimeter (ZDC)~\cite{Col04} provides an estimate of 
the centrality of the collisions.
A more detailed description of the apparatus can be found in \cite{Ban05,Usa05}.
The VT is a radiation-tolerant Si pixel detector, which tracks the charged 
particles produced in the collision (${\rm d}N_{\rm ch}/{\rm d}\eta \sim$ 200 
at midrapidity for central \mbox{In-In} interactions).
By matching the tracks measured in the MS with the corresponding 
tracks in the VT, one can access the kinematical variables of the muons before 
their distortion due to multiple scattering and energy loss fluctuations in 
the hadron absorber~\cite{Sha06,Dav06}. 

The results presented in this letter refer to the full 
\mbox{In-In} data sample collected by NA60, corresponding to $\sim$230 million
events, taken at a beam intensity of 10$^7$s$^{-1}$.  
Only the events having at least one interaction vertex with $\ge$4 tracks 
attached to it are selected for the analysis. 
The distribution of the 
longitudinal coordinate of the vertices allows us to identify the target where
the interaction took place with $\sim$200 $\mu$m accuracy. We require the 
interaction vertex to lie, within this tolerance, in one of the 7 In
sub-targets, 1.5 mm thick each and spaced by 7.5 mm.
Matching of tracks in the VT with the MS tracks is then carried out in 
coordinate and momentum space. The matching efficiency, of the order of 
$\sim$60\% for events where a \jpsi\ has been detected in the MS, 
shows no significant centrality dependence.
In principle, MS tracks could be wrongly matched to VT tracks. 
However, in the \jpsi\ mass region, this effect is negligible 
($<$1\%).

Two kinds of event selection have been applied, corresponding to the two 
different analysis techniques detailed hereafter. In the first, 
the matching of the muon tracks is not performed, retaining in this way a 
larger event sample. 
As a quality check, we perform a cut on the transverse 
distance between the extrapolated MS track at the target position and the beam
axis, weighted by the momentum of the track itself. The level of this cut has 
been set at 10\% of the $\chi^2$ probability for each muon, a value that allows 
us to efficiently reject muons produced downstream of the target.
In the second selection, matching is applied and we require that the tracks 
matched to those detected in the MS originate from the reconstructed vertex or 
from the most upstream one, when more than one is found. In this way, we
reject the small percentage ($\sim$4\%) of events where the \jpsi\ originates 
from a downstream interaction of a nuclear fragment produced in a primary 
\mbox{In-In} collision. 
For both selections, we apply a beam pile-up rejection cut, based on the BT, 
requiring two subsequent ions to be separated in time by at least 12 
ns. In this way we avoid a superposition of signals in the read-out 
gate of the ZDC, that would bias the determination of
the zero-degree energy $E_{\rm ZDC}$.   
Finally, in order to reject events at the edges of the MS acceptance we apply
the kinematical cuts $0 < y_{cms} <1$ and 
$-0.5 < \cos\theta_{CS} < 0.5$, where  $\theta_{CS}$ is the polar angle of the 
muons in the Collins-Soper reference frame. In the end, we are left with 
samples of $\sim$45000~\jpsi\ for the first event selection, and $\sim$29000 
for the second.

\begin{figure}[h]
\centering
\resizebox{0.48\textwidth}{!}
{\includegraphics*[bb=16 8 530 338]{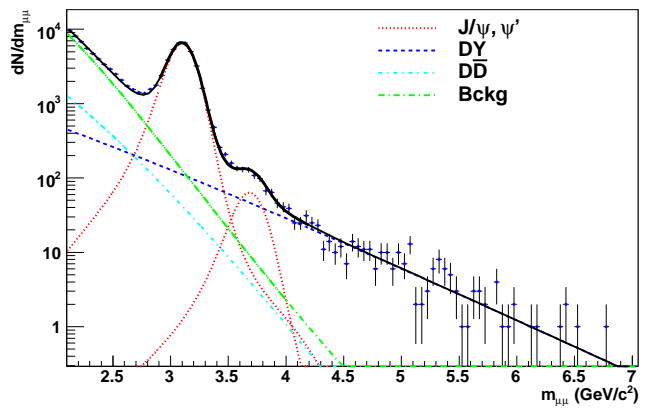}}
\caption{The fit to the $\mu^+\mu^-$ invariant mass spectrum, 
integrated over centrality.}
\label{fig:1}
\end{figure}

The first analysis follows the approach used by the NA38/NA50 
experiments. Namely, the \jpsi\ yield is normalized to the 
corresponding Drell-Yan (DY), a hard process unaffected by final-state 
interactions in the medium~\cite{Ram03}. 
The ratio 
$\sigma_{\rm J/\psi}/\sigma_{\rm DY}$ has the further advantage of being free 
from systematic errors connected with the experimental inefficiencies and the 
integrated luminosity.
It is extracted from a fit of 
the $\mu^+ \mu^-$ invariant mass spectrum 
(in the region $m_{\mu\mu}>$ 2.2 GeV) to a superposition of the  
expected contributions, namely the \jpsi\ and the \psip\ resonances, 
a continuum composed of Drell-Yan events and semi-muonic decays of 
D and $\overline{\rm D}$ mesons, and a combinatorial background from $\pi$ and 
K decays.
The expected mass shapes of the signals and their acceptances are 
evaluated through a Monte-Carlo simulation based on PYTHIA~\cite{Sjo01} 
with GRV94LO~\cite{Glu95} parton distribution functions. 
The combinatorial background has been estimated from the measured sample of 
like-sign pairs and its contribution is negligible. 
Fig.~\ref{fig:1} shows the $\mu^+\mu^-$ 
invariant mass spectrum, together with the result of the fit.

The ratios of cross sections, $\sigma_{\rm J/\psi}/\sigma_{\rm DY}$, 
uncorrected for the \jpsi\ decay branching ratio, are given in 
Table~\ref{tab:1} for three centrality bins corresponding to different values 
of $\langle N_{\rm part}\rangle$, the average number of participant nucleons 
in the bin. $\langle N_{\rm part}\rangle$ has 
been obtained from \ezdc\ using the Glauber model and taking into account the 
smearing induced by the detector resolution. 
The Drell-Yan yield refers to the mass interval 
$2.9 < m_{\mu\mu} < 4.5$ GeV/c$^2$. Its low statistics 
(320 events for $m_{\mu\mu}>$ 4.2 GeV/c$^2$) limits the statistical 
significance of our result and prevents a finer binning 
in centrality. 
\begin{table}
\caption{\label{tab:1} Values of $\sigma_{\rm J/\psi}/\sigma_{\rm DY}$, 
uncorrected for the \jpsi\ decay branching ratio, as a function of centrality.}
\begin{ruledtabular}
\begin{tabular}{ccc}
Centrality bin & $\langle N_{\rm{part}}\rangle$ & 
$\sigma_{\rm J/\psi}/\sigma_{\rm DY}$ \\
\hline
$ E_{\rm ZDC} >$ 11 TeV & 63 & 26.8$\pm$3.2 \\
7 $< E_{\rm ZDC} <$ 11 TeV & 123 & 16.1$\pm$1.6 \\ 
$E_{\rm ZDC} <$ 7 TeV & 175 & 17.8$\pm$1.6 \\
\end{tabular}
\end{ruledtabular}
\end{table}

\begin{figure}[h]
\centering
\resizebox{0.45\textwidth}{!}
{\includegraphics*[bb=18 3 560 366]{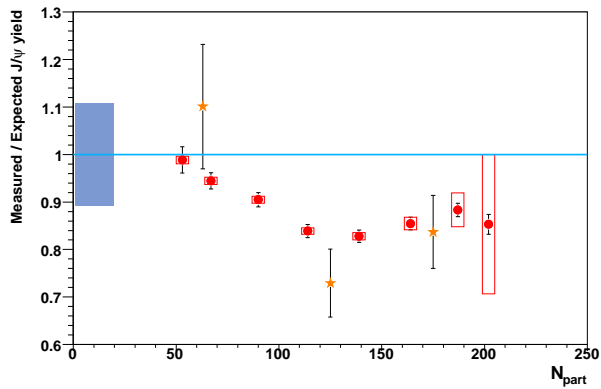}}
\caption{Centrality dependence of the \jpsi\ suppression measured in 
\mbox{In-In} collisions. The stars correspond to the ratio between measured and 
expected $\sigma_{\rm J/\psi}/\sigma_{\rm DY}$, while the circles refer to the 
ratio between the measured \jpsi\ yield and nuclear absorption calculations. 
The box on the left shows the common systematic error, while the boxes around 
the points represent the error related to the centrality determination. 
See text for details.}
\label{fig:2}
\end{figure}

The values shown in Table~\ref{tab:1} indicate that for semi-central and 
central collisions the $\sigma_{\rm J/\psi}/\sigma_{\rm DY}$ ratio is 
significantly lower than for peripheral reactions. In order to understand how 
much of the observed reduction 
is due to cold nuclear 
matter effects, we have calculated, in the frame of the Glauber model, 
$\sigma_{\rm J/\psi}/\sigma_{\rm DY}$ as a function of centrality in a pure 
nuclear absorption scenario. 
This calculation requires as inputs the \jpsi\ absorption cross section in 
cold nuclear matter as well as the ratio of the \jpsi\ and DY elementary 
production cross sections at 158 GeV. Measurements performed by 
NA50~\cite{Ale05,Bor05} in \mbox{p-A} collisions at 450 GeV provide 
$\sigma^{\rm abs}_{\rm J/\psi}$=4.18 $\pm$ 0.35 mb and 
$(\sigma_{\rm J/\psi}/\sigma_{\rm DY})^{pp}_{450}$=57.5 $\pm$ 0.8. The latter 
quantity has been rescaled to 158 GeV/nucleon incident energy, using a 
procedure detailed in Ref.~\cite{Ale05}, obtaining 
$(\sigma_{\rm J/\psi}/\sigma_{\rm DY})^{pp}_{158}$ = 35.7$\pm$3.0.
Fig.~\ref{fig:2} shows the measured $\sigma_{\rm J/\psi}/\sigma_{\rm DY}$, 
divided by the calculated reference and plotted as a function of 
$N_{\rm{part}}$. 
Even if statistical errors are large, a suppression signal of the \jpsi\ beyond 
nuclear absorption can be seen.

\begin{figure}[h]
\centering
\resizebox{0.48\textwidth}{!}
{\includegraphics*[bb=0 0 558 351]{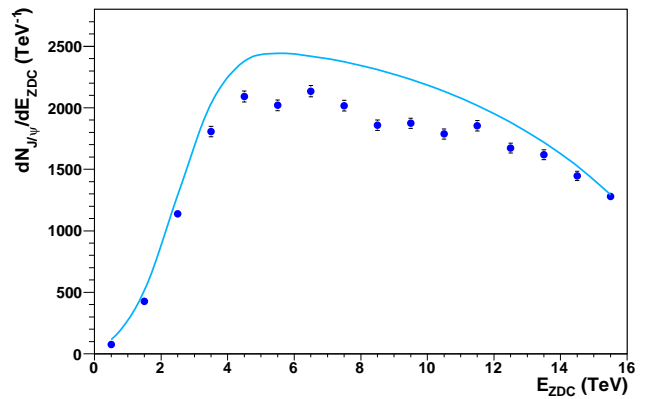}}
\caption{The \jpsi\ $E_{\rm ZDC}$ distribution (circles), compared with 
expectations from nuclear absorption (line).}
\label{fig:3}
\end{figure}

In the second analysis, we directly compare the measured \jpsi\ yield to the 
centrality distribution of \jpsi\ calculated for the case of pure nuclear 
absorption, using the input parameters detailed above. 
In this approach, the measured \jpsi\ yield has been obtained, in 1 TeV \ezdc\ 
bins, by means of a simple fitting procedure that allows us to subtract  
the small amount of Drell-Yan ($<$4\%) and combinatorial background ($<$1\%) 
under the resonance peak.
In Fig.~\ref{fig:3} we 
compare the \jpsi\ distribution with the expectation from nuclear absorption. 
The relative normalization between data and the reference curve is not 
determined a priori; therefore we fix the ratio between data and the nuclear 
absorption curve, integrated over centrality, to the same value 
(0.87 $\pm$ 0.05) as obtained within the previous analysis. It must be noted 
that events where a heavy nuclear fragment
reinteracts in a downstream target have a smaller \ezdc\ value, 
since the fragment reinteraction removes some nucleons that would have 
otherwise reached the ZDC. 
The measured \jpsi\ centrality distribution has been corrected for the small
bias ($<$2\%) induced by this effect.
Finally, because of small inefficiencies of the BT, our data sample could be 
contaminated by high-\ezdc\ events, where a 
non-interacting pile-up ion is superimposed to the interacting one. A 
Monte-Carlo simulation shows that this effect is negligible in our analysis
domain, i.e. \ezdc\ $< 16$ TeV (corresponding to 
$N_{\rm part} > 50$).

Fig.~\ref{fig:2} shows the ratio between the measured and expected \jpsi\ 
yield, rebinned in order to further reduce statistical fluctuations. Since 
statistical errors are very small ($<$2\%), this analysis requires a careful 
estimate of systematic errors. We find that they are connected with the
determination of the shape and normalization of the nuclear absorption 
reference, and with the calculation of $N_{\rm part}$ starting from \ezdc. In 
particular, the uncertainties on 
$(\sigma_{\rm J/\psi}/\sigma_{\rm DY})^{pp}_{158}$ and 
$\sigma^{\rm abs}_{\rm J/\psi}$ give an 8\% and 4\% systematic error on the 
normalization of the absorption curve, respectively.
We then have a 6\% error, originating from the centrality 
integrated value of $\sigma_{\rm J/\psi}/\sigma_{\rm DY}$ used for the
normalization. Concerning centrality determination, by varying within 
errors the input parameters used in the Glauber model, we get a negligible 
influence on the nuclear absorption reference, except for very 
central events (\ezdc$<$3 TeV), where the size of the effect is $\sim$12\%. 
Furthermore, the ZDC does not measure only spectator 
nucleons, but also a small amount of energy released by forward secondary 
particles emitted in the acceptance of the calorimeter 
($\eta > 6.3$). This contribution, important only for central collisions, is 
taken into account when calculating $N_{\rm part}$ from \ezdc. By 
conservatively assuming a 10\% uncertainty on this quantity we get, for events 
with \ezdc$<$3 TeV, a 9\% error on the absorption curve. For more 
peripheral events the effect is negligible. 
Combining the various 
sources in quadrature, we end up with a $\sim$11\% systematic error, 
independent of centrality. 
On top of that, the most central bins are affected by a further, sizeable 
systematic error relatively to the others. 
It must be noted that the systematic error plotted in Fig.~\ref{fig:2}, 
except for the fraction due to the 6\% normalization error quoted above, 
also affects the results on $\sigma_{\rm J/\psi}/\sigma_{\rm DY}$.

\begin{figure}[h]
\centering
\resizebox{0.45\textwidth}{!}
{\includegraphics*[bb=18 3 560 344]{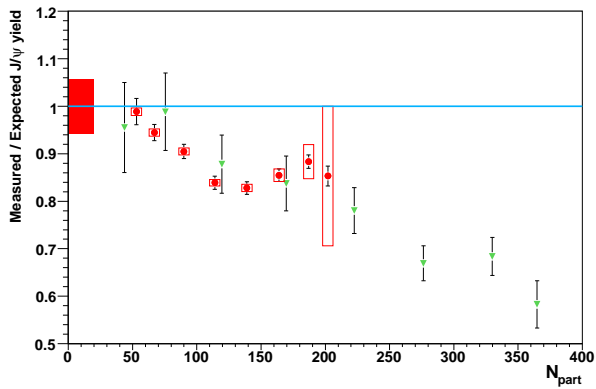}}
\caption{Comparison between the \mbox{In-In} (NA60, circles) and \mbox{Pb-Pb} 
(NA50, triangles) suppression patterns. The box on the left shows the 6\% 
systematic error related to the normalization procedure of the \mbox{In-In} 
points.}
\label{fig:4}
\end{figure}

The results obtained by NA60 in \mbox{In-In} collisions show that in the 
region $50 < N_{\rm part} < 100$ an anomalous suppression of the \jpsi\ sets 
in. Taking into account the $N_{\rm part}$ smearing due to the \ezdc\ 
resolution, the effect seen is compatible with the occurrence of a $\sim$15\% 
drop of the \jpsi\ yield at $N_{\rm part} \sim 80$, followed by a more or less
constant behaviour. When expressed in terms of the Bjorken energy density, the
onset of the anomalous suppression roughly corresponds to 1.5 GeV/fm$^3$ 
(using $\tau_0$=1 fm/c and the VENUS~\cite{Wer93} estimate for the charged
multiplicity as a function of centrality). In Fig.~\ref{fig:4} we compare our 
result with the \jpsi\ suppression pattern 
obtained by NA50 in \mbox{Pb-Pb} collisions~\cite{Ale05}. The systematic errors 
on the determination of the nuclear absorption reference from the \mbox{p-A} 
data sample amount to $\pm$9\% and are not shown in this comparison plot since 
they affect \mbox{Pb-Pb} and \mbox{In-In} results in a similar way.
Within errors, the two patterns look compatible in the $N_{\rm part}$  
region explored by both systems, indicating that $N_{\rm part}$ might be, 
at SPS energy, a scaling variable for the anomalous suppression. A 
detailed investigation of the scaling properties of \jpsi\ suppression as a 
function of several centrality variables  would give valuable insights into 
the origin of the observed effect. However, a meaningful comparison would 
require \mbox{Pb-Pb} results with error bars similar to the ones obtained for 
the \mbox{In-In} analysis. 

Several theoretical predictions, tuned on already available 
\mbox{Pb-Pb} results from NA50, were formulated for \jpsi\ suppression in 
\mbox{In-In} collisions~\cite{Dig04,Cap05,Rap05}. We find that none of them is 
able to quantitatively reproduce the suppression pattern measured by 
NA60~\cite{Sco07}. Recent results from the PHENIX
Collaboration~\cite{Ada06} have shown that also in \mbox{Au-Au} collisions at 
$\sqrt{s}$=200 GeV/nucleon the \jpsi\ is suppressed beyond nuclear
absorption, and that the suppression is larger at forward rapidities.
A coherent interpretation of the results at SPS and RHIC energies is now
mandatory in order to understand the physics mechanisms affecting charmonia 
in a dense partonic/hadronic environment.
The results obtained by NA60 represent the most accurate measurement of \jpsi\ 
suppression in nuclear collisions available today and are a key element to 
strictly constrain theoretical models. Further studies on the $y$ and 
$p_{\rm T}$ dependence of the \jpsi\ suppression are underway and will be the 
subject of future publications.

In summary, we have measured \jpsi\ suppression in \mbox{In-In} collisions at 
158 GeV/nucleon. Comparing the \jpsi\ centrality distribution with the 
expectation from a pure nuclear absorption scenario, we find an anomalous 
suppression that sets in for $N_{\rm part} \sim 80$, and saturates for more 
central events. The statistical errors are negligible (of the order of 2\%). 
Most of the systematic errors are centrality independent and therefore do not 
affect the measured shape of the \jpsi\ suppression pattern. None of the 
existing theoretical models, tuned on the measured \jpsi\ suppression in 
\mbox{Pb-Pb} collisions, is able to quantitatively reproduce the results shown 
in this letter.


\begin{acknowledgments}
The YerPhI group was supported by the C. Gulbenkian Foundation, 
Lisbon and the Swiss Fund Kidagan.
\end{acknowledgments}


\end{document}